\documentclass[twocolumn,aps,pra,showpacs]{revtex4}
\usepackage[T1]{fontenc}
\usepackage[latin9]{inputenc}
\usepackage{amsmath}
\usepackage{amssymb}
\usepackage{graphicx}
\usepackage{esint}

\makeatletter
\@ifundefined{textcolor}{}
{%
 \definecolor{BLACK}{gray}{0}
 \definecolor{WHITE}{gray}{1}
 \definecolor{RED}{rgb}{1,0,0}
 \definecolor{GREEN}{rgb}{0,1,0}
 \definecolor{BLUE}{rgb}{0,0,1}
 \definecolor{CYAN}{cmyk}{1,0,0,0}
 \definecolor{MAGENTA}{cmyk}{0,1,0,0}
 \definecolor{YELLOW}{cmyk}{0,0,1,0}
}

\makeatother

\begin{document}

\title{Collective modes of a one-dimensional trapped atomic Bose gas at
finite temperatures}

\author{Hui Hu$^{1}$, Gao Xianlong$^{2}$, and Xia-Ji Liu$^{1}$, }

\email{xiajiliu@swin.edu.au}

\affiliation{$^{1}$Centre for Quantum and Optical Science, Swinburne University
of Technology, Melbourne 3122, Australia}

\affiliation{$^{2}$Department of Physics, Zhejiang Normal University, Jinhua
321004, China}

\date{\today}
\begin{abstract}
We theoretically investigate collective modes of a one-dimensional
(1D) interacting Bose gas in harmonic traps at finite temperatures,
by using a variational approach and local density approximation. We
find that the temperature dependence of collective mode frequencies
is notably different in the weakly and strongly interacting regimes.
Therefore, the experimental measurement of collective modes could
provide a sensitive probe for different quantum phases of a 1D trapped
Bose gas, realized by tuning the interatomic interaction strength
and temperature. Our prediction on the temperature dependence of the
breathing mode frequency is in good qualitative agreement with an
earlier experimental measurement for a weakly interacting 1D Bose
gas of rubidium-87 atoms in harmonic traps {[}Moritz \textit{et al}.,
Phys. Rev. Lett. \textbf{91}, 250402 (2003){]}. 
\end{abstract}


\maketitle

\section{Introduction}

One-dimensional (1D) systems interacting via a delta-function pairwise
interparticle interaction are of fundamental importance to many-body
physics \cite{Cazalilla2011,Guan2013}. Due to the geometric confinement
in one dimension, the intrinsic strong correlation between particles
may give rise to a number of unusual phenomena, such as effective
fermionization, spin-charge separation, prethermalization and non-trivial
quench dynamics. The recently realized 1D Bose gas of neutral atoms
provides unique opportunities for quantitatively testing and understanding
of these unusual phenomena \cite{Moritz2003,Kinoshita2004,Paredes2004}.
In addition to having the advantage of high degree of control, a 1D
atomic Bose gas can be described by the integrable Lieb-Liniger model
at arbitrary interaction strengths and finite temperatures \cite{Lieb1963,Yang1969},
which is also exactly solvable with powerful numerical techniques
\cite{Schollwock2005} or fundamental mapping theorems \cite{Girardeau1960}.
To date, various physical properties of a 1D atomic Bose gas have
been investigated both theoretically and experimentally, including
the equation of state \cite{Amerongen2008,Jacqmin2011,Vogler2013},
pair correlations \cite{Kheruntsyan2003,Kheruntsyan2005,Kinoshita2005,Esteve2006},
the momentum distribution \cite{Paredes2004,Davis2012} and quench
dynamics \cite{Cheneau2012}. 

The purpose of this paper is to provide a systematic theoretical study
of collective density oscillations of a 1D trapped Bose gas at finite
temperatures. Low-energy collective density modes of a many-particle
system are known to give valuable information about its underlying
physics, at both zero temperature and finite temperatures. In three
dimensions, the measurements of collective density oscillations have
verified the superfluid hydrodynamics at the crossover from Bardeen-Cooper-Schrieffer
superfluids to Bose-Einstein condensates \cite{Hu2004,Altmeyer2007},
and have revealed the second sound propagation in strongly interacting
unitary Fermi gases \cite{Sidorenkov2013}. In two dimensions, such
measurements may be used to examine the scale invariance satisfied
by interatomic interactions \cite{Pitaevskii1997,Vogt2012}. For a
1D Bose gas in harmonic traps with trapping frequency $\omega_{z}$,
collective modes at zero temperature have been extensively investigated
and the breathing mode frequency has been shown to increase from $\sqrt{3}\omega_{z}$
to $2\omega_{z}$ \cite{Moritz2003,Menotti2002,Fuchs2003,Haller2009},
when the cloud crosses from the weakly interacting quasicondensate
regime to strongly correlated Tonks-Girardeau (TG) regime. However,
surprisingly, less is known about collective modes at finite temperatures,
although the first measurement was provided more than ten years ago
\cite{Moritz2003}. 

In this paper, by using a variational approach \cite{Taylor2008,Taylor2009}
together with the exact Yang-Yang equation of state \cite{Yang1969},
we predict the mode frequency of low-lying collective modes of a 1D
trapped Bose gas, at arbitrary interaction strengths and finite temperatures.
We show that the temperature dependence of mode frequencies provide
a useful means to characterize the different quantum regimes of the
1D Bose cloud. Our exact results allow for a stringent parameter-free
comparison with future experimental measurements.

The rest of paper is organized as follows. In the next section, we
briefly describe the exact Yang-Yang thermodynamics of a 1D trapped
Bose gas within the local density approximation. In Sec. III, we outline
the variational approach for the calculation of mode frequencies of
low-lying collective modes. In Sec. IV, we first discuss our theoretical
results (see, for example, Fig. \ref{fig3}) and then compare the
prediction of the breathing mode frequency with the existing experimental
measurement (see Fig. \ref{fig7}). Finally, in Sec. V we draw our
conclusion.

\section{Yang-Yang thermodynamics of a 1D trapped Bose gas}

We consider an atomic gas of $N$ bosons interacting via a pairwise
$\delta$-function potential in one dimension in harmonic traps. In
the first quantization representation, it can be described by the
following Lieb-Liniger Hamiltonian \cite{Lieb1963},
\begin{equation}
\mathcal{H}_{LL}=\sum_{i}\left[-\frac{\hbar^{2}}{2m}\frac{\partial^{2}}{\partial z_{i}^{2}}+V_{ext}\left(z_{i}\right)\right]+\sum_{i<j}g_{1D}\delta\left(z_{i}-z_{j}\right),
\end{equation}
where $m$ is the mass of bosons, 
\begin{equation}
V_{ext}\left(z\right)=\frac{1}{2}m\omega_{z}^{2}z^{2}
\end{equation}
is the harmonic trapping potential with frequency $\omega_{z}$, and
$g_{1D}$ ($>0$) is the only coupling constant used to characterize
the short-range repulsive interaction between bosons. In cold-atom
laboratory, the 1D configuration can now be routinely realized for
atomic gases confined in highly elongated traps or two-dimensional
optical lattices, provided that the transverse degrees of freedom
are frozen out \cite{Moritz2003,Kinoshita2004,Paredes2004}.

In the absence of harmonic traps ($V_{ext}=0$), the Lieb-Liniger
Hamiltonian is integrable. In 1960s, it was exactly solved by Lieb
and Liniger at zero temperature \cite{Lieb1963} and by Yang and Yang
at finite temperatures \cite{Yang1969}. By numerically solving two
coupled integral equations for the density of quasimomenta and the
excitation spectrum \cite{Yang1969}, all the equations of state,
including the number density $n$, entropy $s$, energy $E$ and pressure
$P$, can be determined as functions of the chemical potential $\mu$
and temperature $T$, at a given coupling constant $g_{1D}$. Different
quantum phases of the system can be conveniently characterized by
using the pair correlation function \cite{Kheruntsyan2003,Kheruntsyan2005},
\begin{equation}
g^{\left(2\right)}\left(z,z\right)=-\frac{1}{n^{2}}\left(\frac{\partial P}{\partial g_{1D}}\right)_{\mu,T}.
\end{equation}
For an extensive discussion of the Yang-Yang integral equations and
finite-temperature pair correlations, we refer to the previous work
by Kheruntsyan and his colleagues \cite{Kheruntsyan2005}. 

With the harmonic trapping potential $V_{ext}(z)\neq0$, it is convenient
to use the local density approximation, which amounts to describing
the system locally as a uniform gas with chemical potential equal
to the local effective chemical potential,
\begin{equation}
\mu\left(z\right)=\mu-\frac{1}{2}m\omega_{z}^{2}z^{2},
\end{equation}
where $\mu$ is the global equilibrium chemical potential at the trap
center and is determined by the number equation $N=\int_{-\infty}^{+\infty}dzn(z)$.
All the thermodynamic variables can then be obtained by integrating
the corresponding local quantity over the whole atomic cloud. In particular,
the trap-averaged pair correlation is given by,
\begin{equation}
\overline{g^{(2)}}=\frac{\int_{-\infty}^{+\infty}dzg^{\left(2\right)}\left(z,z\right)n^{2}\left(z\right)}{\int_{-\infty}^{+\infty}dzn^{2}\left(z\right)},
\end{equation}
which can be used to characterize quantum phases in the presence of
harmonic traps. We note that the local density approximation is applicable
for a large system. For the case of a 1D atomic gas, with typical
experimental parameters, a number of atoms $N\sim100$ is often sufficiently
large to ensure the validity of local density approximation \cite{Cazalilla2011,Guan2013}.

\begin{figure}
\begin{centering}
\includegraphics[clip,width=0.48\textwidth]{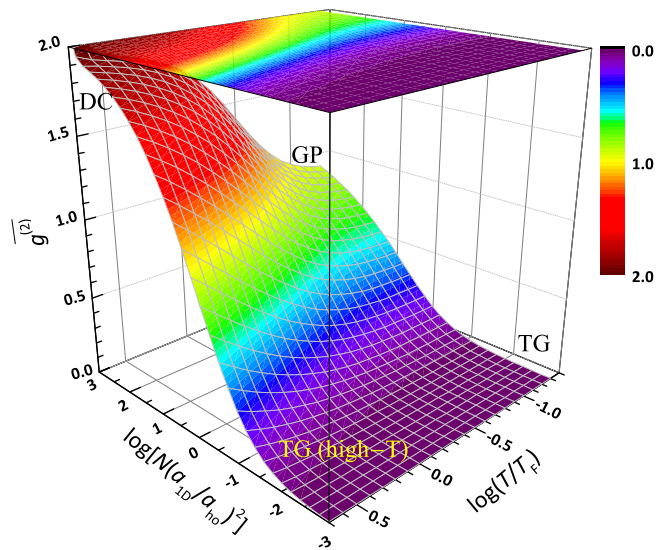} 
\par\end{centering}

\caption{(Color online) Phase diagram of a trapped 1D interacting Bose gas
at finite temperatures. According to the different behavior of the
averaged pair correlation function $\overline{g^{(2)}}$, the system
can be in the Tonks-Girardeau, Gross-Pitaevskii, or decoherent classical
(DC) regime \cite{Kheruntsyan2005}. The transition between different
regimes is continuous.}

\label{fig1} 
\end{figure}

In Fig. \ref{fig1}, we show a contour plot of the averaged pair correlation
function $\overline{g^{(2)}}$ as functions of the dimensionless interaction
parameter $N(a_{1D}/a_{ho})^{2}$ and reduced temperature $T/T_{F}$,
in the logarithmic scale. Here, according to the standard notation
for a trapped 1D gas, we have used the 1D scattering length 
\begin{equation}
a_{1D}=\frac{2\hbar^{2}}{mg_{1D}}
\end{equation}
and $a_{ho}=\sqrt{\hbar/(m\omega_{z})}$ \cite{Hu2007,Liu2007,Liu2008}.
We have also defined the Fermi temperature $T_{F}=N\hbar\omega_{z}$
of an ideal, non-interacting Fermi gas with the same number of atoms
as the unit of temperature \cite{Hu2007,Liu2007,Liu2008}. At low
temperatures, in the weakly-interacting Gross-Pitaevskii (GP) and
strongly-interacting TG regimes, the trapped Bose gas is characterized
by $\overline{g^{(2)}}\simeq1$ and $\overline{g^{(2)}}\simeq0$,
respectively \cite{Kheruntsyan2003}. With increasing temperature,
the weakly interacting Bose cloud quickly turns into a classical Boltzmann
gas, which has $\overline{g^{(2)}}\simeq2$ \cite{Kheruntsyan2005}.
In contrast, the pair correlation function of the strongly interacting
Bose gas is less affected by temperature, although the system will
finally enter a decoherent (quantum) regime at sufficiently high temperatures,
in which the pair correlation will also increase to $\overline{g^{(2)}}\simeq2$
\cite{Kheruntsyan2005}.

\begin{figure}
\begin{centering}
\includegraphics[clip,width=0.48\textwidth]{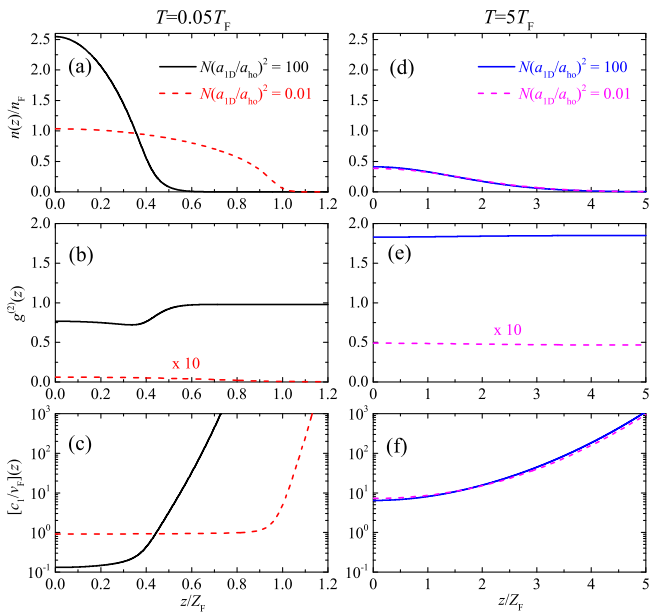} 
\par\end{centering}

\caption{(Color online) Density profile (a, d), local pair correlation (b,
e), and local sound velocity (c, f) in different regimes of temperature
and interaction strength. The left and right panels show the results
at low ($T=0.05T_{F}$) and high temperatures ($T=5T_{F}$), respectively.
The solid and dashed lines refer to the results at the weak and strong
interaction strengths: $N(a_{1D}/a_{ho})^{2}=10^{2}$ (solid lines)
and $N(a_{1D}/a_{ho})^{2}=10^{-2}$ (dashed lines). The density and
position are taken in units of the Fermi density and Fermi radius
of an ideal, spinless Fermi gas with the same number of atoms, $n_{F}=\sqrt{2N}/(\pi a_{ho})$
and $Z_{F}=\sqrt{2N}a_{ho}$, respectively. In the panels (c) and
(f), $v_{F}(z)=\pi\hbar n(z)/m$ is the local Fermi velocity.}

\label{fig2} 
\end{figure}

In Fig. \ref{fig2}, we present the typical density profiles (a, d)
and the local pair correlation functions (b, e) in different quantum
phases, as a function of the axial coordinate normalized to the Fermi
radius $Z_{F}=\sqrt{2N}a_{ho}$. At low temperature ($T=0.05T_{F}$),
the density profiles and local pair correlation are significantly
different in the weak and strong coupling regimes. At high temperature
($T=5T_{F}$), while the density profiles of weakly interacting and
strongly interacting Bose gases are more or less the same, their local
pair correlations differs by about a factor of 40 in magnitude.

\section{Theoretical framework}

At zero temperature, collective density oscillations are governed
by the following hydrodynamic equation for the density fluctuation
$\delta n(z)$,
\begin{equation}
m\omega^{2}\delta n\left(z\right)+\frac{\partial}{\partial z}\left\{ n\left(z\right)\frac{\partial}{\partial z}\left[\left(\frac{\partial\mu}{\partial n}\right)\delta n\left(z\right)\right]\right\} =0,\label{eq:eqdnT0}
\end{equation}
which has been solved by using sum-rule approach \cite{Menotti2002}
or by deriving analytical or quasi-analytical solutions \cite{Fuchs2003}.
At finite temperatures, it is difficult to find a closed equation
for the density fluctuation. Instead, collective modes are better
described in terms of a displacement field $u(z)$,
\begin{equation}
m\left(\omega^{2}-\omega_{z}^{2}\right)nu\left(z\right)+\frac{\partial}{\partial z}\left[n\left(\frac{\partial P}{\partial n}\right)_{\bar{s}}\frac{\partial u\left(z\right)}{\partial z}\right]=0,\label{eq:equz}
\end{equation}
where the derivative of the pressure with respect to density is taken
at a fixed entropy per particle $\bar{s}=s/n$. The above equation
- first derived by Griffin, Wu and Stringari for a collisionally hydrodynamic
Bose gas above $T_{c}$ \cite{Griffin97} - can be derived from the
standard Landau two-fluid hydrodynamic equations \cite{Taylor2009}.
By using the thermodynamic identity 
\begin{equation}
\left(\frac{\partial P}{\partial n}\right)_{\bar{s}}=n\left(\frac{\partial\mu}{\partial n}\right)
\end{equation}
that is valid at zero temperature and applying the definition of the
density fluctuation 
\begin{equation}
\delta n\left(z\right)=-\frac{\partial}{\partial z}\left[nu\left(z\right)\right],
\end{equation}
it is easy to check that at zero temperature Eq. (\ref{eq:equz})
indeed reduces to the equation for density fluctuation Eq. (\ref{eq:eqdnT0}).

In the absence of harmonic traps, the solution of Eq. (\ref{eq:equz})
is a plane wave of wave vector $q$ with dispersion $\omega=cq$,
where the sound velocity $c$ is given by,
\begin{equation}
c=\sqrt{\frac{1}{m}\left(\frac{\partial P}{\partial n}\right)_{\bar{s}}}.
\end{equation}
To numerically calculate $(\partial P/\partial n)_{\bar{s}}$, we
note that 
\begin{equation}
\left(\frac{\partial P}{\partial n}\right)_{\bar{s}}=\frac{\left(P_{\mu}s_{T}-P_{T}s_{\mu}\right)-\bar{s}\left(P_{\mu}n_{T}-P_{T}n_{\mu}\right)}{\left(n_{\mu}s_{T}-n_{T}s_{\mu}\right)},
\end{equation}
where we have used the notations such as $P_{\mu}\equiv(\partial P/\partial\mu)_{T}$
and $s_{T}\equiv(\partial s/\partial T)_{\mu}$. All these first-order
derivatives can be obtained numerically by solving the Yang-Yang integral
equations \cite{note1}.

In Figs. \ref{fig2}(c) and \ref{fig2}(f), we report the local sound
velocity in units of the local Fermi velocity $v_{F}(z)$ for different
quantum phases of a 1D interacting Bose gas. At low temperatures ($T=0.05T_{F}$),
sound velocity is dramatically different in the weakly and strongly
interacting regimes. However, the difference is washed out quickly
by increasing temperature. As a result, far above the degenerate temperature
($T=5T_{F}$), sound velocity becomes nearly the same, regardless
of the strength of interatomic interactions.

\subsection{Variational approach}

In the presence of harmonic traps, the finite-temperature hydrodynamic
equation for the displacement field can be solved by using a variational
approach \cite{Taylor2008,Taylor2009}. For this purpose, we reformulate
it using Hamilton's variational principle and introduce a hydrodynamic
action, 
\begin{equation}
S=\int dz\left[m\left(\omega^{2}-\omega_{z}^{2}\right)nu^{2}-n\left(\frac{\partial P}{\partial n}\right)_{\bar{s}}\left(\frac{\partial u}{\partial z}\right)^{2}\right].\label{action_firstsound}
\end{equation}
All the collective modes can be obtained by minimizing $S$. Therefore,
we consider the following variational (polynomial) ansatz for the
displacement field $u\left(z\right)$: 
\begin{equation}
u\left(z\right)=\sum_{i=0}^{N_{p}-1}A_{i}z^{i}.
\end{equation}
The precision of our variational calculations can be improved with
increasing $N_{p}$, which is the number of the expansion basis \{$A_{0},A_{1},\cdots,A_{N_{p}-1}$\}.
Inserting the variational displacement field into the action (\ref{action_firstsound}),
after some straightforward algebra we find that, 
\begin{equation}
S=\left[A_{0}^{*},...,A_{i}^{*},...\right]{\cal S}\left(\omega\right)\left[A_{0},...,A_{j},...\right]^{T},
\end{equation}
where the superscript ``$T$'' stands for the transpose of vector
and ${\cal S}\left(\omega\right)$ is a $N_{p}\times N_{p}$ matrix
with elements 
\begin{equation}
{\cal S}_{ij}\left(\omega\right)=\omega^{2}M_{ij}-K_{ij}.
\end{equation}
Here, 
\begin{equation}
M_{ij}=m\int dzn\left(z\right)z^{i+j}
\end{equation}
and
\begin{equation}
K_{ij}=m\omega_{z}^{2}\int dzn\left(z\right)z^{i+j}+ij\int dzn\left(\frac{\partial P}{\partial n}\right)_{\bar{s}}z^{i+j-2}
\end{equation}
are the weighted mass moments and the spring constants, respectively.
It is readily seen that the minimization of the action $S$ is equivalent
to solving 
\begin{equation}
\mathcal{S}\left(\omega\right)\left[A_{0},...,A_{j},...\right]^{T}=0
\end{equation}
or 
\begin{equation}
\det{\cal S}\left(\omega\right)=0.
\end{equation}

\subsection{Dipole mode and breathing mode}

It is easy to see that, the dipole mode - for which the displacement
field $u(z)$ is a constant - is decoupled from other modes, as a
result of the fact that $K_{ij}=\omega_{z}^{2}M_{ij}$ if $i=0$ or
$j=0$. This is consistent with Eq. (\ref{eq:equz}), in which the
second term involving the equation of state vanishes for a constant
displacement field. Hence, the dipole mode is an exact solution of
the hydrodynamic equation, with a temperature independent frequency
$\omega_{z}$. In the literature, this is known as Kohn's theorem,
which holds for any Galileo transformation invariant systems.

The next mode, the so-called breathing mode, is not an exact solution
of the hydrodynamic equation and its mode frequency relies on the
equation of state of the system, through the speed of sound $(\partial P/\partial n)_{\bar{s}}$.
However, at zero temperature it is known that the displacement $u(z)=A_{1}z$
provides an excellent variational ansatz, which essentially becomes
exact in the deep weakly interacting GP or strongly interacting TG
regimes \cite{Menotti2002,Fuchs2003}. As a result, the breathing
mode frequency $\omega_{B}$ can be well approximated by, 
\begin{equation}
\omega_{B}^{2}\simeq\frac{K_{11}}{M_{11}}=\omega_{z}^{2}+\frac{\int_{-\infty}^{+\infty}dz\left[n\left(\partial P/\partial n\right)_{\bar{s}}\right]}{m\int_{-\infty}^{+\infty}dzn\left(z\right)z^{2}}.\label{wb2sumrule}
\end{equation}
Indeed, at zero temperature this sum-rule approach works extremely
well \cite{Menotti2002}. The resulting breathing mode frequency differs
very slightly from the exact theory, with a relative error at the
order of $0.1\%$ \cite{Fuchs2003}.

\section{Results and discussions}

Our variational approach converges very quickly with increasing the
number of the variational parameter $N_{p}$. The results presented
below are calculated with $N_{p}=8$, for any given temperature $T/T_{F}$
or dimensionless interaction parameter $N(a_{1D}/a_{ho})^{2}$.

\subsection{Asymptotic behavior of mode frequencies in some limits}

Before discussing our numerical results, it is useful to briefly summarize
the known asymptotic behavior of mode frequencies in the low and high
temperature limits. At zero temperature, the local chemical potential
takes the form $\mu(r)\propto n(r)$ in the weakly interacting GP
limit and $\mu(r)\propto n^{2}(r)$ in the strongly interacting TG
limit, respectively. As a result, the hydrodynamic equation at zero
temperature Eq. (\ref{eq:eqdnT0}) admits exact polynomial solutions
with mode frequency \cite{Menotti2002,Fuchs2003}
\begin{equation}
\omega_{n}=\sqrt{\frac{n\left(n+1\right)}{2}}\omega_{z}
\end{equation}
in the GP limit and with frequency \cite{Minguzzi2001} 
\begin{equation}
\omega_{n}=n\omega_{z}
\end{equation}
in the TG limit, where the integer $n=1,2,\cdots$ is the index of
collective modes. At large temperatures, the equation of state of
the Bose gas is well-approximated by the Boltzmann distribution and
the density profile takes a gaussian distribution. In this case, the
finite temperature hydrodynamic equation Eq. (\ref{eq:equz}) can
again admit polynomial solutions. The mode frequency is given by,
\begin{equation}
\omega_{n}=\sqrt{3n-2}\omega_{z}.
\end{equation}
It is clear that in all cases $\omega_{n=1}=\omega_{z}$ as required
by the Kohn theorem. The breathing mode is the first mode that has
non-trivial temperature dependence. It takes the mode frequency $\omega_{B}=\sqrt{3}\omega_{z}$
in the low-temperature mean-field regime and $\omega_{B}=2\omega_{z}$
in either the strongly interacting limit or high-temperature regime.

\begin{figure}
\begin{centering}
\includegraphics[clip,width=0.45\textwidth]{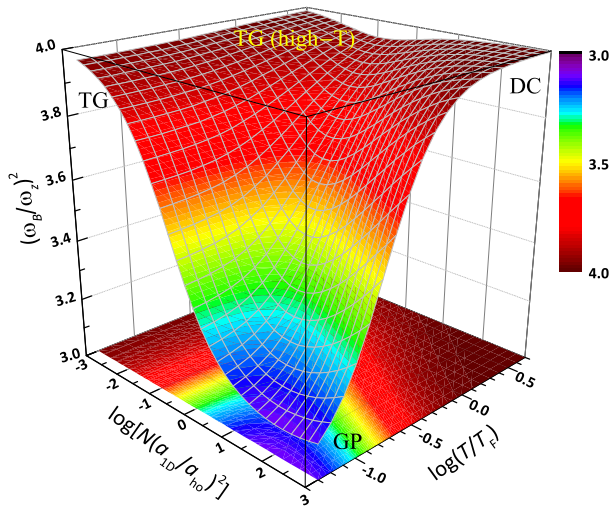} 
\par\end{centering}

\caption{(Color online) Contour plot of the square of the breathing mode frequency
$(\omega_{B}/\omega_{z})^{2}$ in different regimes. With increasing
temperature or interaction strength, the breathing mode frequency
increases from $\sqrt{3}\omega_{z}$ to $2\omega_{z}$. }

\label{fig3} 
\end{figure}

\begin{figure}
\begin{centering}
\includegraphics[clip,width=0.45\textwidth]{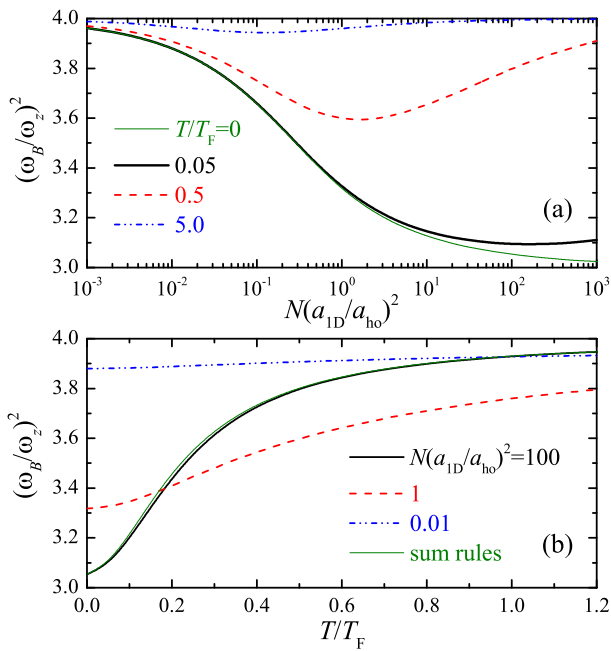} 
\par\end{centering}

\caption{(Color online) The square of the breathing mode frequency $(\omega_{B}/\omega_{z})^{2}$
as a function of interaction strength $N(a_{1D}/a_{ho})^{2}$ (a)
or as a function of temperature $T/T_{F}$ (b). For a weakly interacting
Bose gas at $N(a_{1D}/a_{ho})^{2}=100$, the breathing mode frequency
obtained by the sum-rules approach is also shown by a thin line in
(b).}

\label{fig4} 
\end{figure}

\subsection{Breathing mode frequency}

In Fig. \ref{fig3}, we present a three-dimensional contour plot of
the breathing mode frequency. In Figs. \ref{fig4}(a) and \ref{fig4}(b),
we plot respectively the coupling constant and temperature dependence
of the breathing mode frequency in greater detail. Here, in both figures
we show $(\omega_{B}/\omega_{z})^{2}$ rather than $\omega_{B}/\omega_{z}$,
following the sum-rule convention used in the previous theoretical
studies \cite{Menotti2002}.

For the coupling constant dependence, see for example Fig. \ref{fig4}(a),
the mode frequency at zero temperature decreases monotonically from
the asymptotic strongly interacting value $2\omega_{z}$ at $N(a_{1D}/a_{ho})^{2}\ll1$
to the weakly interacting value $\sqrt{3}\omega_{z}$ at $N(a_{1D}/a_{ho})^{2}\gg1$,
in agreement with the analytic analysis mentioned earlier. At finite
temperatures, however, this monotonic decrease no longer persists.
At sufficiently weak coupling, we find that the mode frequency will
finally turn up due to a nonzero temperature, whatever how small it
is. As a result, a broad minimum appears in the mode frequency, with
its position shifting to the strong coupling regime when temperature
increases.

\begin{figure}
\begin{centering}
\includegraphics[clip,width=0.48\textwidth]{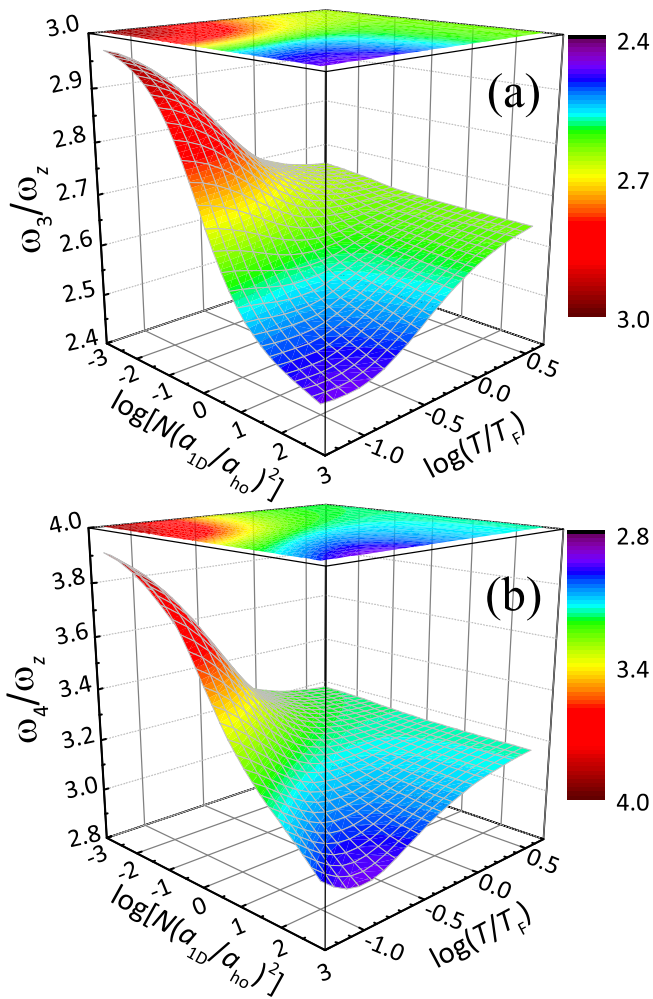} 
\par\end{centering}

\caption{(Color online) Contour plot of the mode frequency for the third (upper
panel, $\omega_{3}/\omega_{z}$) and fourth (lower panel, $\omega_{4}/\omega_{z}$)
collective modes.}

\label{fig5} 
\end{figure}

For the temperature dependence as shown in Fig. \ref{fig4}(b), we
find instead that the mode frequency always increases with increasing
temperature at arbitrary interaction strength. In the strongly interacting
regime, the trend of increasing becomes very weak, as the zero temperature
mode frequency itself shifts to the same value of $2\omega_{z}$ as
in the high temperature limit. We note that, at finite temperatures
the sum-rule result Eq. (\ref{wb2sumrule}) still works excellently
well. At the coupling constant $N(a_{1D}/a_{ho})^{2}=100$, the sum-rule
result and the full variational prediction for $(\omega_{B}/\omega_{z})^{2}$
differ in relative by $1\%$ at most, indicating that the breathing
mode is indeed well-approximated by the displacement field $u(z)\propto z$.

\begin{figure}
\begin{centering}
\includegraphics[clip,width=0.48\textwidth]{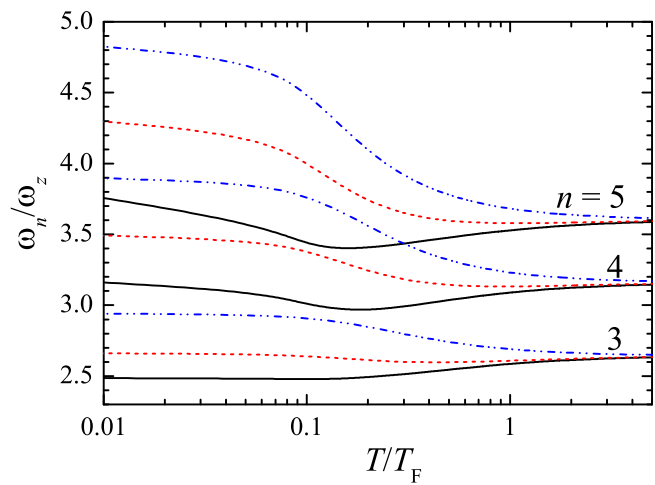} 
\par\end{centering}

\caption{(Color online) Temperature dependence of the mode frequency for higher-order
modes ($n=3,4,5$) at three different interaction strengths: $N(a_{1D}/a_{ho})^{2}=100$
(solid line), $1$ (dashed line), and $0.01$ (dot-dashed line).}

\label{fig6} 
\end{figure}

\subsection{Higher mode frequency}

We now consider the mode frequency of higher collective modes. Although
these modes are technically more difficult to excite than the breathing
mode, their excitations in quasi-1D configuration has been recently
demonstrated for a unitary Fermi gas \cite{Tey2013}. In Fig. \ref{fig5},
we report the contour plot of the third and fourth compressional modes.
The temperature dependence of the frequency of higher modes is presented
in Fig. \ref{fig6}. 

The qualitative behavior of the frequency of higher collective modes
is very similar to what we have observed in the breathing mode. The
most remarkable difference is that the mode frequency at zero temperature
may significantly be larger than the frequency at high temperatures.
As a result, the mode frequency no longer increases monotonically
with increasing temperature. Nonetheless, it is clear that all the
mode frequencies could have distinct behavior in different quantum
phases and hence could provide a useful way to characterize the phase
diagram, in addition to the pair correlation characterization.

\begin{figure}
\begin{centering}
\includegraphics[clip,width=0.48\textwidth]{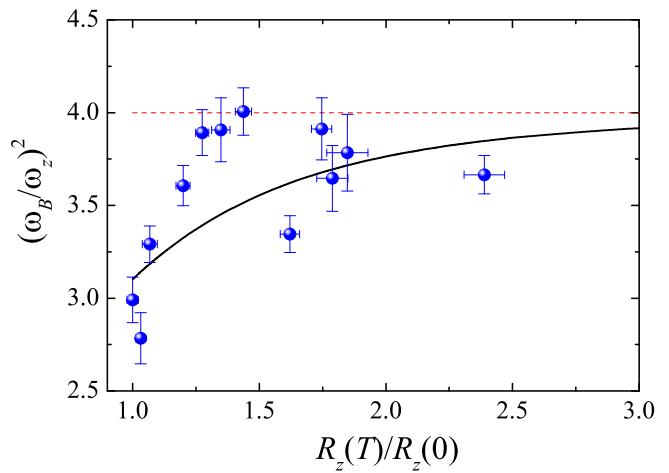} 
\par\end{centering}

\caption{(Color online) Comparison between our theoretical predictions (solid
line) and the experimental data (solid circles with error bars) \cite{Moritz2003},
for the temperature dependence of breathing mode frequency. Experimentally,
the temperature is characterized by the rms cloud size $R_{z}=\sqrt{\left\langle z^{2}\right\rangle }$
measured after 15 ms of time-of-flight \cite{Moritz2003}. In the
experiment, the coupling constant $N(a_{1D}/a_{ho})^{2}\simeq18$.}

\label{fig7} 
\end{figure}

\subsection{Comparison with the experiment}

Experimentally, the temperature dependence of the breathing mode frequency
of a 1D interacting Bose gas at a particular interaction strength
was measured by the Esslinger team over ten years ago. The temperature
of the atomic cloud is tuned by varying the hold time prior to excitations
of the breathing mode and is indirectly characterized by measuring
the axial width of the cloud after $15$ milliseconds of time-of-flight
expansion \cite{Moritz2003}. Therefore, it is not possible to directly
compare our theoretical result of the breathing mode frequency with
the experimental data. 

In Fig. \ref{fig7}, we compare the theory and experiment by assuming
that the normalized cloud size, given by the ratio between the rms
axial width at temperature $T$ and at zero temperature $R_{z}(T)/R_{z}(0)$,
is not affected by the time-of-flight expansion. The agreement between
theory and experiment is qualitative good, without any free adjustable
parameters. The discrepancy is due to the assumption of invariant
ratio $R_{z}(T)/R_{z}(0)$ during the time-of-flight expansion as
well as the fact that the Bose cloud prior to excitations may not
be fully in thermal equilibrium \cite{Moritz2003}. Refined experimental
measurements are required, in order to fully understand the temperature
dependence of the breathing mode frequency.

\section{Conclusions}

In conclusion, we have investigated the finite-temperature collective
modes of a one-dimensional interacting Bose gas in harmonic traps,
which are described by the Landau hydrodynamic equation. We have used
Hamilton's variational principle and have derived a variational action
for solving the hydrodynamic equation. By taking a polynomial variational
ansatz, we have accurately calculated the frequency of collective
modes at arbitrary temperature and interaction strength.

The mode frequency of collective modes, particularly the breathing
mode frequency, is found to have pronounced temperature dependence
in different quantum phases in the weak and strong coupling regimes.
As a result, experimental measurement of collective modes could provide
a sensitive probe of the phase diagram of a one-dimensional trapped
Bose gas, complementing the proposed characterization of pair-correlation
functions at finite temperatures \cite{Kheruntsyan2003,Kheruntsyan2005}.
We have shown that our theoretical prediction on the temperature dependence
of the breathing mode frequency is in good qualitative agreement with
an earlier experimental measurement for a weakly interacting 1D Bose
gas of rubidium-87 atoms in harmonic traps \cite{Moritz2003}. Quantitative
test of our theory could be obtained in the near future with high-precision
measurements for collective modes. 
\begin{acknowledgments}
HH and XJL were supported by the ARC Discovery Projects (Grant Nos.
FT130100815, DP140103231 and DP140100637) and NFRP-China (Grant No.
2011CB921502). GX was supported by the Zhejiang Provincial Natural
Science Foundation under Grant No. R6110175.

\textit{Note added}. --- In completing our numerical calculations,
we become aware of a recent experimental measurement of the breathing
mode in 1D atomic Bose gases reported by Isabelle Bouchoule and co-workers
\cite{Fang2014}. In their preprint, the breathing mode frequency
has also been theoretically calculated by directly solving the coupled
hydrodynamic equation for small density fluctuations.\end{acknowledgments}


\begin{thebibliography}{10}
\bibitem{note1}As an important check for numerical accuracy, we use
the thermodynamic relations, $P_{\mu}=n$, $P_{T}=s$, and $s_{\mu}=n_{T}$.
\end{thebibliography}
\end{document}